\documentclass{pasa}
\usepackage{textcomp}
\usepackage{gensymb}
\usepackage{soul}
\usepackage{amsmath}
\usepackage{amssymb}
\usepackage{booktabs}
\usepackage{graphicx}

\usepackage[plainpages=false, colorlinks=true, anchorcolor=blue, linkcolor=blue, citecolor=blue, bookmarks=false]{hyperref}
\usepackage{float}
\restylefloat{table}
\restylefloat{figure}
\usepackage[pdftex]{color}
\usepackage{listings}
\usepackage{enumitem}
\usepackage{cuted}
\usepackage[bottom]{footmisc}

\definecolor{dkgreen}{rgb}{0,0.6,0}
\definecolor{gray}{rgb}{0.5,0.5,0.5}
\definecolor{mauve}{rgb}{0.58,0,0.82}

\definecolor{lightgray}{rgb}{0.9,0.9,0.9}
\definecolor{darkgreen}{rgb}{0,0.4,0}

\DeclareMathOperator*{\argmin}{arg\,min}
\newcommand{\mnras}{Mon. Not. R. Astron. Soc.}
\newcommand{\apj}{Astrophys. J.}
\newcommand{\apjs}{Astrophys. J. Suppl. Ser.}
\newcommand{\prd}{Physical Review D.}
\newcommand{\aj}{Astron. J.}
\newcommand{\apjl}{Astrophys. J. Lett.}
\newcommand{\araa}{Annual Reviews of Astronomy and Astrophysics}

\newcommand{\forloop}[5][1]%
{%
\setcounter{#2}{#3}%
\ifthenelse{#4}%
	{%
	#5%
	\addtocounter{#2}{#1}%
	\forloop[#1]{#2}{\value{#2}}{#4}{#5}%
	}%
	{%
	}%
}%

\newcommand{\ctbd}[1]{}

\title[ADVI]{Variational Inference as an alternative to MCMC for parameter estimation and model selection}

\author{Geetakrishnasai Gunapati$^1$,
\thanks{\url{cs19mtech11019@iith.ac.in}}
Anirudh Jain$^2$,
P.K. Srijith$^1$,
Shantanu Desai$^3$
\affil{$^{1}$ Dept. of Computer Science and Engineering, IIT Hyderabad, Kandi, Telangana 502285, India}
\affil{$^{2}$ Dept. of Computer Science, Aalto University, Espoo 02150; Finland}
\affil{$^{3}$ Dept. of Physics, IIT Hyderabad, Kandi, Telangana 502285, India
}
}
\jid{PASA}
\doi{10.1017/pas.\the\year.xxx}
\jyear{\the\year}
\begin{document}

\begin{frontmatter}
\maketitle

\begin{abstract}

Most applications of Bayesian Inference for parameter estimation and model selection in astrophysics involve the use of Monte Carlo techniques such as Markov Chain Monte Carlo (MCMC) and nested sampling. However, these techniques are  time consuming and  their convergence to the posterior could be difficult to determine.   In this work, we advocate Variational inference  as an alternative to solve the above  problems, and demonstrate its usefulness for parameter estimation and model selection in Astrophysics. Variational inference converts the inference problem into an optimization problem by approximating the posterior from a known family of distributions and using Kullback-Leibler divergence to characterize the difference. It takes advantage of fast optimization techniques, which make it ideal to deal with large datasets and makes it trivial to parallelize on a multicore platform. We also derive a new approximate evidence estimation  based on variational posterior, and importance sampling technique called posterior weighted importance sampling for the calculation of evidence (PWISE), which is  useful  to perform Bayesian model selection.
As a proof of principle, we apply variational inference to five different problems in astrophysics, where Monte Carlo techniques were previously used. These include assessment of significance of annual modulation in the COSINE-100 dark matter experiment, measuring exoplanet orbital parameters from radial velocity data, tests of periodicities in measurements of Newton's constant $G$, assessing the significance of a turnover in the spectral lag data of GRB 160625B and estimating the mass of a galaxy cluster using weak gravitational lensing. We find that variational inference is much faster than MCMC and nested sampling techniques for most of these problems while providing competitive results. All our analysis codes have been made publicly available.
\end{abstract}

\end{frontmatter}

\section{Introduction}
 Markov Chain Monte Carlo (MCMC) is the most common method for inference, and for sampling multi-modal probability distributions~\citep{hastings1970monte,gelfand1990sampling,Sharma,2017arXiv171006068H,SpeagleMCMC}. 
Following  the rapid rise in the  usage of Bayesian analysis in astronomy, MCMC (and nested sampling) techniques  are now widely used (starting with ~\citealt{Saha}) for a variety of problems   ranging from parameter estimation, model comparison, evaluating model goodness of fit,  to forecasting for future experiments. This is because it is usually not possible to analytically calculate the multi-dimensional integrals needed for computing the Bayesian posteriors or evidence, and  the numerical evaluation of these integrals can easily get intractable. Also, almost all numerical optimization techniques run into problems while maximizing the Bayesian posterior, when the total number free parameters gets large. For this reason, there has been an unprecedented surge in the usage of Monte Carlo techniques in astrophysics.
However, MCMC techniques are not tied only to Bayesian methods. They have also been used in frequentist analysis, for sampling complex multi-dimensional likelihood needed for parameter estimation~\citep{Meszaros}. That said, the ubiquity of MCMC methods in Astronomy has been driven by the increasing usage of  Bayesian methods.
Applications of MCMC to a whole slew of astrophysical problems have been recently reviewed in  ~\citep{Sharma}. Although a large number of MCMC sampling methods have been used,  the most widely used MCMC sampler in Astrophysics is {\tt Emcee}~\citep{emcee}. Bayesian model comparison is usually done using Nested sampling~\citep{Skilling}, which is also a Monte Carlo based technique. A large number of packages have been used in Astrophysics for carrying out Bayesian model comparison using Nested Sampling techniques,  such as {\tt MultiNest}~\citep{multinest}, {\tt Nestle}~\footnote{\url{http://kylebarbary.com/nestle/}}, {\tt dynesty}~\citep{dynesty} etc. These techniques are however  computationally expensive.

Although, MCMC  has evolved into one of the most important tools for Bayesian inference~\citep{robert2011short}, there are problems for which we cannot easily use this approach, especially in the case of large datasets or models with high dimensionality.  Variational inference~\citep{jordan1999introduction} provides a good alternative approach for approximate Bayesian inference and has been the subject of considerable research recently~\citep{blei2017variational}. It provides an approximate posterior for Bayesian inference faster than simple MCMC by solving an optimization problem. \citet{pmlr-v33-ranganath14} and \citet{kucukelbir2016automatic} compare the convergence rates for variational inference against other sampling algorithms. They both show that variational inference convergences much faster in lesser number of iterations, even when the Metropolis-Hastings algorithm doesn't converge for the same problem.

The use of variational inference with deep learning is becoming more widespread in Astrophysics, especially in the areas of image generation and classification. Generating reliable synthetic data that can be used as calibration data for future surveys is an important task, which otherwise is a expensive task. ~\citet{ravanbakhsh2016enabling, Spindler_2020,Bastien_2021} have used conditional variational auto encoder (cVAE) for the task of image generation. ~\citet{ravanbakhsh2016enabling} used cVAE with convolutional layers and adversarial loss to generate galaxy images using galaxy zoo dataset,~\citet{Bastien_2021} used cVAE with fully connected layers for the task of generating synthetic images from radio galaxies. ~\citet{Walmsley_2019} used Bayesian neural networks (BNN) for calculating posterior over image labels, which can provide uncertainties for each label for a given image. This can be converted to traditional deterministic classification by collapsing posterior to corresponding point estimates.

~\citet{Jiang_2021} used BNN for tracing fibrils in the  H$\alpha$ images of the sun. A specific BNN dubbed FibrilNet was used for the  segmentation task i.e the probability of each pixel being a fibril is predicted with a uncertainty, then a fibril fitting algorithm is used on this mask to trace firbils and identify their orientation. A significant number of confirmed exoplanets (about 4000 which is 30\% of all identified exoplanets) have been identified through the validation of false positive cases from non-planet scenarios. ~\citet{Armstrong_2020} used Gaussian process classifier (GPC) for this validation task and showed that their method is much faster than the competing algorithm \textbf{vespa} with comparable results.~\citet{Lin_2021} combined deterministic deep learning classifier CLDNN (it combines CNN and a LSTM) with variational inference to detect events of binary coalescence in observation data of gravitational waves along with uncertainty estimates. This can be used in real time detection of events and the events with high uncertainty can be pushed for further examination rather than accepting or discarding event. ~\citet{moralesalvarez2019scalable} used variational gaussian processes for tackling the problem of crowdsourcing in Glitch detection in LIGO. They show that variational gaussian processes very well compared to other traditional deep learning techniques and also take less time to train.

VI has also been used in the task of parameter estimation. ~\citep{Hort_a_2020} combined BNN with normalizing flows (NF) for estimating astronomical and cosmological parameters from 21cm surveys. ~\citet{gabbard2020bayesian} use cVAE for estimating the source parameters for gravitational wave detection. They show that the estimated parameters are close to the parameters from traditional MCMC algorithms. The significant amount of time taken in this process is training of the cVAE network; it takes about $\mathcal{O}$(1) day. Once trained the network need not be trained again, and the   GW detection parameters can be obtained 6 orders of magnitude faster, when compared to existing techniques.

Few works were done comparing MCMC and VI approached. In the work done by ~\citep{Schlegel}, a generative model for constructing astronomical catalogs using  telescope image datasets was developed using Bayesian inference. They developed two approximate inference procedures using MCMC and variational inference for their statistical model and compared the effectiveness of the methods. The aforementioned paper found that  for the synthetic data generated from their model, MCMC was better in estimating uncertainties, but it was about three orders slower compared when compared to the competing variational inference procedure. Whereas on real data taken from SDSS, the uncertainty estimates in both the  procedures  were far from perfect. In that work, they were successful in applying variational inference to the entire SDSS data, thus demonstrating its feasibility on very large datasets. This technique has also been used in lensing for estimating the uncertainties in parameters through Bayesian neural networks~\cite{Blundell} for the problem of Singular Isothermal Ellipsoid plus external shear and total flux magnification~\citep{PerreaultLevasseur:2017ltk}. Recently,~\citet{Hort_b_2020} used BNN for estimating parameters for cosmic microwave background. They found that VI was 4 orders faster when compared to MCMC with slight compromise in accuracy. They also showed that using output from BNN as initial proposal for Markov chain resulted in higher acceptance rate for Metropolis-Hasting algorithm.

For the purpose of computing Bayesian evidence, needed for model comparison,
~\citep{Bernardo03thevariational} have compared Variational Bayes and Annealed importance sampling (AIS)~\citep{neal2001} for the task of evidence estimation and posterior evaluation. Their results show that Variational Bayes is about 100 times faster when compared to AIS without any significant loss in accuracy. 



In this work, we shall explain how a particular adaptation of variational inference (dubbed ADVI) can supersede Monte Carlo techniques such as MCMC and  nested sampling for parameter estimation and Bayesian model comparison and apply these techniques to five different problems in astrophysics and compare the results to Monte Carlo methods. The outline of this paper is as follows.  In section
~\ref{sec:bayes_mcmc}, we introduce the idea of Bayesian modeling and provide an introduction to MCMC. In section
~\ref{sec:secVI}, we present an overview of the variational inference method. In section~\ref{sec:advi}, we discuss a specific implementation of variational inference called Automatic differentiation variational inference (ADVI).  In section~\ref{sec:param}, we explain how variational inference  can be used for parameter estimation and model comparison. Applications to ancillary problems in astronomy are outlined in section~\ref{sec:examples}. We conclude in section~\ref{sec:conclusions}. The code for all the analyses in this manuscript can also be found on a github link provided at the end of this manuscript.

\section{Overview of Bayesian Modeling and MCMC}
\label{sec:bayes_mcmc}
We first  start with a very brief primer on Bayesian modeling and parameter inference, and then explain how Monte Carlo methods  are applied to these problems. More details on Bayesian methods and their applications in astrophysics are reviewed in ~\citep{Trotta,Sharma,Weller} and references therein.
Bayes Theorem in general terms is given as,
\begin{equation}
\label{eq:bayes}
p(\theta|\textbf{D}) = \frac{ p(\textbf{D}, \theta) }{p(\textbf{D})} = \frac{ p(\textbf{D} |\theta) p(\theta )}{p(\textbf{D})},
\end{equation}

\noindent where $p(\theta)$ is the prior belief on the parameter $\theta$, $p(\textbf{D} |\theta)$ is known as  the likelihood, which models the probability of observing the data $\textbf{D}$ given parameter $\theta$. $p(\theta|\textbf{D})$ called posterior probability, is the conditional probability of $\theta$ given $\textbf{D}$, which can be interpreted as the posterior belief over the parameters after evidence or data $\textbf{D}$ is observed. $p(\textbf{D})$ is termed as the marginal likelihood or model evidence, which is obtained by integrating out $\theta$ from the joint probability distribution $p(\textbf{D}, \theta) $, the numerator term in Eq.~\eqref{eq:bayes}. All the conditional probabilities in Eq.~\eqref{eq:bayes} are implicitly conditioned on the model $m$. Hence the marginal likelihood $p(\textbf{D})$ provides the probability that the model $m$ will generate the data irrespective of its parameter values and is a useful quantity for model selection.   

Bayesian models treat the parameters as a random variable and impose preliminary knowledge about the parameter through the prior. Inference in the Bayesian model amounts to conditioning on the data and computing the posterior $P(\theta|\textbf{D})$. This computation is intractable for models where the prior and likelihood take different functional forms (non-conjugates). In these cases, analytical closed form estimation of the marginal likelihood is also intractable. This has led to the usage of sampling methods to solve for such intractable distributions.

MCMC methods are sampling techniques, which enable us to sample for any unnormalized distribution~\citep{hastings1970monte,gelfand1990sampling,Sharma,2017arXiv171006068H,SpeagleMCMC}. The idea of MCMC algorithms is to construct and sample from a Markov chain whose stationary distribution is  the same as the desired distribution, and use those samples to compute expectations and integrals of required quantities using Monte Carlo integration techniques.  We will briefly introduce the Metropolis-Hastings algorithm (M-H)~\citep{Metropolis,hastings1970monte}, which is the simplest  MCMC algorithm. Although the M-H algorithm is simple, it shares many of the same principles with the newer and more complex MCMC algorithms. M-H algorithm requires a proposal distribution $q(\theta'|\theta)$, which is used to generate parameter samples. Assume the unnormalized posterior distribution over the parameters $\theta$ to be represented as the function $f(\theta)$, i.e. $f(\theta) \propto p(\textbf{D} |\theta)p(\theta)$. The M-H algorithm works as follows.
\begin{itemize}
    \item Assume that $\theta_k$ is the previous sampled point, draw the next sample $\theta'$ from the proposal distribution $q(\theta'|\theta_k)$
    \item Draw a random number r from a uniform distribution between 0 and 1
    \item Accept the sample if $\frac{f(\theta')q(\theta_t|\theta')}{f(\theta_k)q(\theta'|\theta_t}) > r$ ($\theta_{k+1}\leftarrow\theta'$)  else reject the sample ($\theta_{k+1}\leftarrow\theta_k$)
\end{itemize}

When run long enough, the M-H algorithm produces samples from the desired posterior distribution. Although the algorithm is simple, there are many different parameters in the algorithm that are to be tuned to achieve ideal results. One of the important parameters for the algorithm is the number of samples that the algorithm has to run for achieving reliable results. There is nothing called absolute convergence for a MCMC algorithm and one can only rely on heuristics. We can run multiple chains with different initial points and can compare posterior inferences like the mean and variance from both the chains. There are other metrics like autocorrelation time~\citep{Sokal1997} and Gelman–Rubin diagnostic~\citep{gelman1992} which can be used to check for pseudo-convergence of MCMC algorithms.

Choosing a proposal distribution also plays a vital role in the quality of samples that are produced. A proposal distribution that is too narrow can result in accepting all the samples and will take a lot of time covering the entire parameter space, while a proposal distribution that is too wide can result in taking large steps and rejecting most of the samples. For example, consider a Gaussian distribution $\mathcal{N}(0,\sigma)$ as the proposal distribution and $\theta_k$ is the current sample, then the next sample $\theta'$ is calculated as $\theta' \leftarrow \theta_k + \mathcal{N}(0,\sigma)$. The value of $\sigma$ dictates the distance between the two proposal and it is the step size in this case. One can use a simple heuristic like the  acceptance ratio for tuning the step size, high acceptance ratio means that you are accepting all the generated samples and hence has to reduce the step size and vice-versa. The choice of proposal distribution is not problem independent and finding efficient proposal distribution can become increasingly difficult with increase in dimensions of parameter space. 

Initialization like proposal distribution is an input parameter to most of the MCMC algorithms. A badly initialized chain can spend a lot of time in regions of low probability, which can result in a large number iterations for the MCMC algorithm to reach a stationary condition. In such cases we discard a certain number of initial samples from the chain before the stationary condition is reached. This idea is called as burn-in and the length of burn-in depends on each individual problem and initialization. If the proposal distribution is multi-modal, then starting multiple chains with different initializations and comparing the samples will help in identifying if chains have covered all the modes. If different initializations result in different chains, then there is no straight forward method of combining the samples from multiple chains. One has to run a MCMC algorithm for a long time so that each chain can cover all the modes, and produce a representative sample or resort to Nested sampling techniques.

There are many advanced methods like tempering~\citep{Vousden_2015}, which help the MCMC samplers from being stuck at one mode in multi-modal distributions.  Hamiltonian Monte Carlo (HMC)~\citep{betancourt2018conceptual} which uses the gradients of the function $f(\theta)$ for efficient generation of proposals. HMC avoids the random walk sampling approach and hence can be efficient in exploring parameter space even for high dimensional cases. HMC's performance is sensitive to two tunable parameters: the step size $\epsilon$ and the  desired number of steps $L$. If $L$ is too small then HMC ends up exhibiting random walk behaviour which is undesirable, and if $L$ is too high the algorithm can waste a lot of computational power. No-U-Turn Sampler (NUTS)~\citep{hoffman2011nouturn} is an extension to HMC which eliminates the manual tuning of $L$ and calculates the number of steps through a recursive algorithm. Therefore, NUTS is as efficient as HMC if not better in most of the cases and eliminates the need for manual tuning.

Affine invariant ensemble sampling uses multiple random walkers for drawing proposal samples and it significantly outperforms the standard M-H algorithm in drawing independent samples with much lesser autocorrelation time ~\citep{goodman2010,ForemanMackey2013}. Nested sampling~\citep{Feroz_2019} converts the multi-dimensional integration of evidence \textbf{D} into a 1-D integration by mapping likelihood to the corresponding prior volume in the corresponding iso-likelihood contours on a 2-D curve. This 1-D curve integration can be evaluated using trapezoid rule. MCMC methods as seen, may require a lot of tuning and in most cases this tuning can require a deeper mathematical understanding of algorithm being used for achieving desirable results.

Therefore in this work, we study a alternative method for performing Bayesian inference called called as variational inference, which is considerably faster than MCMC techniques and does not suffer from any convergence issues.

\section{Variational Inference}
\label{sec:secVI}

The central idea behind variational inference is to solve an optimization problem by approximating the target probability density. The target probability density could be the Bayesian posterior or the likelihood from frequentist analysis.
The first step is to propose  a family of densities and then to find the member of that family,  which is closest to the target probability density. Kullback-Leibler divergence~\citep{kullback1951} is used as a measure of such proximity. 

For this purpose,  we then posit a family of approximate densities (variational distribution) $\mathcal{Q}$. This is a set of densities over the parameters. It is important to choose an complex enough variational family such that the target distribution lies in it, otherwise the solution obtained will  not be close to the  target probability distribution. Then, we try to find the member of that family $q(\theta) \in \mathcal{Q}$, known as  the variational posterior that minimizes the Kullback-Leibler
($\mathrm{KL}$) divergence to the exact posterior,

\begin{equation}
\label{eq:posteriorApprox} q^*(\theta) = \argmin_{q(\theta) \in \mathcal{Q}} \mathrm{KL}(q(\theta) | | p(\theta | \textbf{D}))
\end{equation}

The $\mathrm{KL}$ divergence is defined as,

\begin{equation}
\label{eq:KLDiv1} \resizebox{0.9\hsize}{!}{$\mathrm{KL} (q(\theta) | | p(\theta | \textbf{D}))= \mathbb{E}_{q(\theta)} [\log q(\theta)] - \mathbb{E}_{q(\theta)}[\log p(\theta | \textbf{D})],$}
\end{equation}

\noindent where all the expectations are with respect to $q(\theta)$. We shall see in Eq.~\eqref{eq:KLDiv2} that  $\mathrm{KL}$ divergence depends on the posterior $\log p(\theta | \textbf{D})$, which is usually intractable to compute.  We can expand the conditional using ~\eqref{eq:bayes} and re-write  $\mathrm{KL}$ divergence as,
{
\small
\begin{eqnarray}
\label{eq:KLDiv2}  \mathrm{KL} (q(\theta) | | p(\theta | \textbf{D})) &=& \mathbb{E}_{q(\theta)}[\log q(\theta)]  + \mathbb{E}_{q(\theta)}[\log  p(\textbf{D})] \nonumber \\
&&-\mathbb{E}_{q(\theta)}[\log  p(\textbf{D}, \theta)] \nonumber \\
&=& \log  p(\textbf{D}) + \mathbb{E}_{q(\theta)}[\log q(\theta)] \nonumber \\
&&-\mathbb{E}_{q(\theta)}[\log  p(\textbf{D}, \theta)] .
\end{eqnarray}
}%


The expected value of the log evidence with respect to the variational posterior is the log evidence term itself, and is independent of the variational distribution. Hence, minimizing the $\mathrm{KL}$ divergence term is equivalent to minimizing the second and third terms in Eq.~\eqref{eq:KLDiv2}. Equivalently, one could estimate the variational posterior by  maximizing the variational lower bound (also known as evidence lower bound or ELBO~\citep{blei2017variational})  with respect to $q(\theta)$. 
\begin{equation}
\label{eq:ELBO1} \mathrm{ELBO}(q(\theta)) = \mathbb{E}_{q(\theta)}[\log p(\textbf{D}, \theta)] - \mathbb{E}_{q(\theta)}[\log q(\theta)] .
\end{equation}
ELBO can be viewed as a lower bound to the evidence term by re-arranging the terms in Eq.~\eqref{eq:KLDiv2}. 
{
\small
\begin{equation}
\label{eq:logevidence}
\log  p(\textbf{D}) =  \mathrm{KL} (q(\theta) | | p(\theta | \textbf{D})) + \mathrm{ELBO}(q(\theta)).
\end{equation}
}%
The KL divergence between any two distributions is a non-negative quantity and hence, $\log  p(\textbf{D}) \geq \mathrm{ELBO}(q(\theta))$. Again, we can see that as the evidence term is independent of the variational distribution, maximizing ELBO will result in minimizing the KL divergence between the variational posterior and the actual posterior. 

\noindent Expanding the joint likelihood in Eq.~\eqref{eq:ELBO1}, the variational lower bound can be rewritten as: 
{
\small
\begin{eqnarray}
\label{eq:ELBO2} \mathrm{ELBO}(q(\theta)) &=& \mathbb{E}_{q(\theta)}[\log p(\textbf{D}| \theta)] - \mathbb{E}_{q(\theta)}[\log q(\theta)]\nonumber \\
&&+ \mathbb{E}_{q(\theta)}[\log p(\theta)] \nonumber \\
&=& \mathbb{E}_{q(\theta)}[\log p(\textbf{D}| \theta)] - \mathrm{KL} (q(\theta) | | p(\theta)).
\end{eqnarray}
}%
 The first term in Eq.~\eqref{eq:ELBO2}, which  can be interpreted as the data fit term,  will result in selecting a variational posterior, which maximizes the likelihood of observing the data. While the second term can be seen as the regularization term, which minimizes the KL divergence between the variational posterior and the prior. Thus, ELBO implicitly regularizes the selection of the variational posterior and trades-off likelihood and prior in arriving at a proper choice for the variational posterior. The log evidence term in Eq.~\eqref{eq:logevidence} and hence the variational lower bound  (ELBO) are implicitly conditioned on the hyper-parameters of the model.  The hyper-parameters can be learned by maximizing the variational lower bound.  Typically, the variational parameters and the hyper-parameters are learned alternatively by  maximizing the variational lower bound. 

Variational inference converts the Bayesian parameter estimation into an optimization problem through the  maximization of the variational lower bound. 
Hence, convergence is guaranteed in variational inference, as is the case of any optimization problem, to a local optimum and if the likelihood is log-concave then to a global optimum. Another important feature of variational inference is that it is trivial to  parallelize. It can handle large datasets with ease without compromising on the model complexity with the use of stochastic variational inference~\citep{hoffman2013stochastic}. In the case of some specific likelihoods and variational families, $\mathrm{ELBO}$ cannot be computed in closed form as the computations of required expectations are intractable. In these settings, either one resorts to model specific algorithms~\citep{jaakkola1996computing,blei2007correlated,braun2010variational} or generic algorithms that require model specific calculations~\citep{knowles2011non,wang2013variational,paisley2012variational}.

Recent advances in variational inference use ``black box'' techniques to avoid model specific lower bound calculations~\citep{pmlr-v33-ranganath14,kingma2013auto,rezende2014stochastic,salimans2014using,titsias2014doubly}. These ideas were leveraged to develop automatic differentiation variational inference techniques (ADVI)~\citep{kucukelbir2016automatic} that works on any model written in the probabilistic programming systems such as  {\tt Stan}~\citep{carpenter2016stan}\footnote{We have used the ADVI implementation  in {\tt PyMC3} for our case studies} or {\tt PyMC3}~\citep{salvatier2016}

\section{Automatic Differentiation Variational Inference}
\label{sec:advi}
Variational inference algorithm requires model specific computations to obtain the variational lower bound. Typically, variational inference requires the manual calculation of a custom optimization objective function by choosing a variational family relevant to the model, computing the objective function and its derivative, and running a gradient-based optimization.

Automatic differentiation variational inference (ADVI)~\citep{kucukelbir2016automatic} automates this by building a ``black-box'' variational inference technique, which takes a probabilistic model and a dataset as inputs and returns posterior inferences about the model's latent variables. ADVI achieves the results by performing the following sequence of steps.

\begin{itemize}
\item ADVI applies a transformation on the latent variables $\theta$ to obtain   real-valued latent variables $\zeta$, where $\zeta = T(\theta)$ and $\zeta \in \mathbb{R}^{dim(\theta)}$. The transformation $T$ ensures that  all the latent variables lie on a real co-ordinate space, and allows ADVI to use the same variational family $q(\zeta; \phi)$ (for e.g. Gaussian where $q(\zeta; \phi) = \mathcal{N}(\zeta;\mu,\Sigma)$)  on all the models. This transformation changes the variational lower bound and the joint likelihood $p(\textbf{D}, \theta)$ is written in terms of $\zeta$ as $p(\textbf{D}, \zeta) = p(\textbf{D}, T^{-1}(\zeta)) |J_{T^{-1}}(\zeta)|$, where $|\cdot|$ represents the determinant. Here, $J_{T^{-1}}(\zeta)$ is the Jacobian of the inverse of the transformation $T$. The variational lower bound takes the following form under this 
transformation.

{
\footnotesize
\begin{align}
    ELBO(q(\zeta; \phi))=&\mathbb{E}_{q(\zeta; \phi)} [\log p(\textbf{D}, T^{-1}(\zeta)) +log |J_{T^{-1}}(\zeta)|] \nonumber \\
                        &-\mathbb{E}_{q(\zeta; \phi)} (\log q(\zeta; \phi)).
\end{align}
}%

\item The variational objective (ELBO) as a function of the variational parameters $\phi$ (for instance mean $\mu$ and covariance $\Sigma$ of a Gaussian) can be optimized using gradient ascent. 

However, the calculation of gradients of ELBO with respect to the  variational parameters is generally intractable.
To push the gradients inside the expectation, ADVI applies elliptical standardization. Consider a transformation $S_{\phi}$, which absorbs the variational parameters $\phi$ and converts the non-standard Gaussian $\zeta$ into a standard Gaussian $\eta$, $\eta =  S_{\phi} (\zeta)$. For instance, $\eta = L^{-1}(\zeta - \mu)$, where $L$ is the Cholesky factor for the covariance $\Sigma$.  The expectation in the variational lower bound can be written in terms of the standard Gaussian  $\displaystyle q(\eta) = \mathcal{N}(\eta;0,I)$ and the variational lower bound becomes:

{
\footnotesize
\begin{align}
\label{eq:elbo_zeta}
ELBO(q(\zeta; \phi)) =& \mathbb{E}_{\mathcal{N}(\eta;0,I)} [\log p(\textbf{D}, T^{-1}(S_{\phi}^{-1}(\eta))) \nonumber \\ 
&+ \log |J_{T^{-1}}(S_{\phi}^{-1}(\eta))|]  + \mathbb{H}(q(\zeta; \phi)).
\end{align}
}%


\item The entropy term in Eq.~\eqref{eq:elbo_zeta} is problem independent and its gradient can be evaluated in closed form for a Gaussian distribution. Therefore, its gradients are evaluated before hand and are used for all the problems. The variational lower bound Eq.~\eqref{eq:elbo_zeta} has expectations independent of 
$\zeta$, and hence the gradient of ELBO with respect to $\phi$ can be calculated by pushing the gradient inside the expectations. 

{
\footnotesize
\begin{align}
\nabla_{\phi} ELBO(q(\zeta; \phi)) =& \mathbb{E}_{\mathcal{N}(\eta;0,I)} [\{\nabla_{\theta} \log p(\textbf{D},\theta) \nabla_{\zeta} T^{-1} \nonumber \\
&+ \nabla_{\zeta} log |J_{T^{-1}}(\zeta)\}\nabla_{\phi}S_{\phi}^{-1}(\eta)] \nonumber \\
&+\nabla_{\phi}\mathbb{H}(q(\zeta; \phi)) .
\end{align}
}%


The gradients inside the expectations are computed using automatic differentiation, while the expectation with respect to the standard Gaussian is computed using Monte Carlo sampling. The values of $\zeta = S_{\phi}^{-1}(\eta)$ and $\theta = T^{-1}(S_{\phi}^{-1}(\eta))$ at corresponding $\eta$ are calculated and substituted while evaluating the expectation.
\end{itemize}

\section{Parameter Estimation  and Bayesian Model Selection}
\label{sec:param}
Once we have the  approximate posterior, we can draw samples from the variational posterior over the parameters. 
Unlike in MCMC, the number of samples required is not an input to the optimization and it does not affect the training time of variational inference. We  can find a point estimate of the parameters using the mean (or median)  of the samples from  the variational posterior. In certain cases, we consider the variational distribution family to be parameterized by the mean, and we learn the variational posterior by maximizing the variational lower bound with respect to the mean. In these cases, we can directly make use of the mean rather than sampling from the variational posterior. The errors and marginalized credible intervals for the parameters can be obtained by passing the samples from ADVI (similar to MCMC) to the {\tt corner} module~\citep{corner}
or similar packages such as {\tt ChainConsumer}~\citep{Hinton2016} or {\tt GetDist}~\citep{getdist}. 

\label{sec:modelcomp}
A major challenge in statistical modeling is choosing a proper model,  which generates the observations.  In a Bayesian setting, one could use a posterior probability over the models in choosing the right model. Consider two models $M_1$ and $M_2$ with a prior probability over  them denoted by $p(M_1)$ and $p(M_2)$. The probability of these models generating the observations irrespective of the parameter values is given by the evidence (marginal likelihood)  $p(\textbf{D}~|~M_1)$ and   $p(\textbf{D}~|~M_2)$. Combining the prior and the likelihood, one could obtain the posterior over the models $p(M_1~|~\textbf{D})$ and $p(M_2~|~\textbf{D})$.

As discussed earlier, the evidence term is   computed by evaluating the integral over the parameter likelihood and prior:
\begin{equation}
\label{evidenceequ}
p(\textbf{D}~|~M) = \int p(\textbf{D}~|~\theta, M) p(\theta~|~M) d\theta .
\end{equation}
This is independent of  $\theta$ and represents a normalization constant associated with the posterior.  The evidence term provides the probability of generating the data by some model $M$. It implicitly penalizes models with high complexity through the Bayesian Occam's Razor~\citep{mackay1992,murphy13}.   Complex models  (models with large number of parameters) will be able to generate a wider set of  observations  but with a lower probability for each set of observation, since  $p(\textbf{D}~|~M)$ over observation sets should sum to unity. While simpler models will be able to generate only a fewer set of observations with a higher probability to each set of observations. For given set of observations $\textbf{D}$, one could choose an appropriate model based on the complexity involved in generating $\textbf{D}$.  If $\textbf{D}$  is simple, we will choose a simple model.  Simple models will be able to provide high likelihood values  $p(\textbf{D}~|~\theta, M)$ for a large number of parameter values $\theta$, and the prior value $p(\theta~|~M)$ also takes higher values as the parameter space is small.  When the model complexity increases the prior over the parameters $p(\theta~|~M)$ takes a lower value.  Also, a complex model will give a high likelihood value only for a small number of  parameters. For a large number of parameter values, it will not be able to model simple data sets. 

\subsection{Posterior weighted importance sampling for evidence}
\label{sec:approxevidence}
The evidence term $p(\textbf{D}~|~M)$ is intractable for non-conjugate cases, and variational inference provides a lower bound to the evidence term (ELBO), which acts as a proxy to the evidence. The tightness of the ELBO bound depends on how close the approximate posterior  is to the actual posterior.  ELBO provides a good proxy for the evidence only when the variational posterior  is the same as the actual posterior. If the variational approximation assumed is not close to the actual posterior, the bound can be very large and hence using ELBO for model comparison might not be always correct.   In this work, we derive an approximation to Bayesian evidence based on the variational posterior and the importance sampling technique.

Monte Carlo integration technique allows us to approximate Eq.~\eqref{evidenceequ} by replacing the integral with a sum over samples taken from $p(\theta)$. 
\begin{equation}
\label{evidenceque_1}
p(\textbf{D}~|~M) =  \sum_{\theta_i} p(\textbf{D}~|~\theta_i, M) .
\end{equation}
This approximation generally results in a good estimate for the expectation but can require a large number of samples in some cases. Consider a scenario where the likelihood is small in regions where $p(\theta)$ is large, and the  likelihood is large where $p(\theta)$ is small. In such a scenario, the approximation is dominated by regions of low likelihood and can require large number of samples from $p(\theta)$ to achieve the desired estimate.
Importance sampling provides a methodology for efficient sampling for such scenarios. In importance sampling, we choose a proposal distribution and use the samples from the proposal distribution for evaluating the expectation in the Eq. ~\eqref{evidenceequ}.
\begin{eqnarray}
p(\textbf{D}~|~M) &=& \int \frac{p(\textbf{D}~|~\theta, M) p(\theta~|~M)}{q(\theta)}   q(\theta) d\theta . \label{importancesampling} \\
&=& \sum_{\theta_i} \frac{p(\textbf{D}~|~\theta_i, M) p(\theta_i~|~M)}{q(\theta_i)}, \label{approximateevidence}
\end{eqnarray}
where ${\theta}_i$ denote the  samples from the proposal distribution. The quantities $\frac{p({\theta}_i)}{q({\theta}_i)}$ are known as importance weights and these importance weights compensate for the bias introduced because of sampling from $q(\theta)$ instead of $p(\theta)$. It can be easily seen that a proposal distribution should have a large value whenever the product of the likelihood and the prior is large and a small value whenever the product is small. From Eq.~\eqref{eq:bayes} we can see that the posterior is equal to the product of likelihood and prior divided by a normalizing constant and hence is a perfect choice for a proposal distribution. Since the posterior distribution is unknown and is approximated by the variational distribution, we can use the variational distribution as the proposal distribution. We propose to use Eq.~\eqref{approximateevidence}  to compute the  approximate evidence term with  $q(\theta)$ as the variational approximation to the posterior learnt by maximizing ELBO. We call this approximate quantity as posterior weighted importance sampling for evidence (PWISE) and this will be used as a proxy to the evidence (or marginal likelihood) for performing Bayesian model comparison. 

\section{Applications to Astrophysical problems}
\label{sec:examples}
As a proof of principle, we now apply ADVI to five different  problems  from astronomy, particle astrophysics, and gravitation, where MCMC and nested sampling techniques were previously used for parameter estimation and model comparison.  We discuss in detail the  ELBO derivation for one of these problems, namely the COSINE-100 dark matter  experiment, in section ~\ref{subsec:cosine-100}.  We also compare the computational costs using ADVI over MCMC and nested sampling techniques. In this work, we use the  {\tt PyMC3} python package for all our ADVI experiments and {\tt PyMC3} or {\tt emcee} python packages for our MCMC experiments. We also use {\tt nestle} or {\tt dynesty} packages to calculate evidence and compare with our approximate evidence calculation using PWISE.

Previously, ~\citet{Cameron2019} had compared AIS and nested sampling and showed that nested sampling outperforms AIS in many cases with much shorter run time. Although other sampling techniques such as Gaussianized Bridge Sampling~\citep{Seljak19}, proximal nested sampling~\citep{Cai21}, stepping stone algorithm~\citep{Maturana}, diffuse nested sampling~\citep{Brewer}, Adaptive Annealed Importance Sampling~\citep{Liu14} have  been investigated, Nested sampling is most widely used because of the ready availability of packages such as {\tt Dynesty} and {\tt Nestle}. Hence for model selection problems, we check if Nested samling and approximate evidence lead to the same qualitative conclusion using Jeffreys scale.

\subsection{Assessment of significance of annual modulation in cosine-100 data} \label{subsec:cosine-100}
Weakly Interacting Massive Particles (WIMP) are elementary particles beyond the Standard Model of Particle Physics that are hypothesized as dark matter candidates~\citep{Desai04}. Over the past few decades many experiments have been carried out  to detect WIMPs, and out of all of these, only DAMA/LIBRA has identified annual modulations, which show all the correct  characteristics of being generated by WIMP particle interactions~\citep{Dama18}. This result however has been ruled out by many other direct detection experiments. However all these experiements used a target material different than DAMA/LIBRA.
The COSINE-100 experiment  dark matter experiment~\citep{Cosine}   is the first experiment  with target material,  which is a replica of the DAMA/LIBRA  target, and therefore can be used to verify the claims of annual modulation of DAMA/LIBRA using an independent detector target. This experiment has recently started taking data and released its first results about two years ago~\citep{Cosine}.
An independent analysis of this data using Bayesian model comparison methods was carried out in \citep{Krishak_2019}. The COSINE-100 experiment uses data from five different crystals.  The event rate for each of  these crystals is given by:
\begin{equation}
R = C + p_0\exp{(\frac{-\ln{2} \cdot t}{p_1})}+A\cos{\omega(t-t_0)} .
\label{eq:event_rates}
\end{equation}

The last term in Eq.~\eqref{eq:event_rates} corresponds to the annual modulation caused by the WIMP particle interactions~\citep{Freese}. We do a model selection  between two hypothesis: viz., that the data from the crystals consist of the cosine term (H1), versus   without the  cosine term (H2). For this purpose, the data of all the five crystals is fit simultaneously using the  same values for the cosine parameters across all crystals, and crystal specific values for the remaining background-only parameters.

Before we move on to model comparison, we explain  the process involved  in  variational inference and  the lower bound derivation for this problem.  This will provide a deeper theoretical understanding of variational inference and also serve as a motivation for using automatic differentiation variational inference. As discussed in Section ~\ref{sec:secVI}, we first need to posit a family of variational distributions $\mathcal{Q}$ that approximate the posterior distribution. Let us approximate the variational family as a Gaussian distribution with diagonal variance i.e. $q_{\phi}(\theta) = \mathcal{N}(\mu,\Sigma)$. For this particular problem, the likelihood  $P(D|\theta)$ is a Gaussian with mean given by the event rate described in Eq.~\eqref{eq:event_rates} and standard deviation given by the errors in the data. The priors $P(\theta)$ used for all the parameters are uniformly distributed. More details of the analysis and choice of priors can be found in  ~\citep{Krishak_2019}.

\begin{eqnarray*}
\label{eq:cosine_like_prior}
q(\theta) &=& \prod_{i \in {(C,p_0,p_1,A,\omega,t_0)}}  \frac{1}{\sqrt{2\pi\sigma_{i}^2}} \exp(-\frac{(\theta_i-\mu_i)^2}{2\sigma_{i}^2}). \\
p(\theta) &=& \prod_{i \in {(C,p_0,p_1,A,\omega,t_0)}} \frac{1}{max_i - min_i}. \\
p(D|\theta) &=& \prod_{i} \frac{1}{\sqrt{2\pi\sigma_{i}^2}} \exp(-\frac{(r-R_i)^2}{2\sigma_{i}^2}) , 
\end{eqnarray*}
where $\mu_i$ and $\sigma_i$ are the variational parameters (denoted by $\phi$) and $T_i = p_0\exp{(\frac{-\ln{2} \cdot t_i}{p_1})}+A\cos{\omega(t_i-t_0)}$. The variational parameters ($\phi$) are then estimated through evidence lower bound (ELBO) maximization. The ELBO for the cosine problem is given in the equation Eq.~\eqref{eq:elbo_cosine}, and  we will simplify the equation for one chosen latent variable `C' for brevity.


{\small
\begin{align}
\label{eq:elbo_cosine}
\mathrm{ELBO} =& \mathbb{E}_{q(C,p_0,p_1,A,\omega,t_0)}[\log p(\textbf{D}| C,p_0,p_1,A,\omega,t_0)] \nonumber \\
&- \mathrm{KL} (q(C,p_0,p_1,A,\omega,t_0) | | p(C,p_0,p_1,A,\omega,t_0)) \nonumber  \\
=& \mathbb{E}_{q(\theta)}\bigg[ \mathbb{E}_{q(C)}\bigg[\log p(\textbf{D}|\theta)] \log\frac{p(C)}{q(C)}\bigg]\bigg] \nonumber \\
&-\mathrm{KL}\big(q(\theta) | | p(\theta)\big) ,
\end{align}
}%
where $(p_0,p_1,A,\omega,t_0)$ are the latent variables $\theta$.  Eq.~\eqref{eq:elbo_C} shows the final equation for ELBO for the latent variable `C' after substituting for the aforementioned likelihood, prior, and variational distribution. For a detailed derivation of Eq.~\eqref{eq:elbo_C}, please refer to the Appendix.

{\footnotesize
\begin{align}
\label{eq:elbo_C}
\mathrm{ELBO} =& \frac{1}{2} + \log \frac{B\sqrt{2\pi\sigma_{C}^{2}}}{C_{max}-C_{min}} \nonumber \\
&-\sum_i \frac{1}{2\sigma_i^2}\bigg(\mathbb{E}_{q(p_0,p_1,A,\omega,t_0)}\bigg[(r_i-T_i-\mu_C)^2 \bigg]+ \sigma_C^2\bigg)\nonumber \\
&-\mathrm{KL}(q(p_0,p_1,A,\omega,t_0) | | p(p_0,p_1,A,\omega,t_0))
\end{align}
}%
where $\mu_C$ and $\sigma_C$ are the variational parameters describing the posterior over "C". The log term (second term) in Eq.~\eqref{eq:elbo_C} is the result of KL divergence between the variational distribution and the prior distribution. This  acts as a regularization term, which will prevent $\sigma_C$ (variance of variational posterior)  from going to zero during the maximization of ELBO (due to the third term, which is negative).  Consequently, the variational posterior learnt by maximizing ELBO will be a well formed distribution, with probability density not only around the mean but over a larger region covering the posterior. We can calculate the  gradients of the ELBO with respect to the  variational parameters ($\phi$) and use stochastic gradient decent for estimating $\phi$.

The problem of choosing a suitable variational family $\mathcal{Q}$ is not always easy. Consider the above case where the variational distribution is the Gaussian distribution. The prior for ``C'' is a uniform distribution between 0 and 400, which implies  that  the mean of the posterior distribution $\mu$ should be a positive value. But there is no explicit condition present in Eq.~\eqref{eq:elbo_C} that constrains the $\mu$ to take only positive values after optimization. Therefore the choice of the variational family $\mathcal{Q}$ depends on each individual problem and involves solving a complex constrained optimization problem.

ADVI mitigates the above problems by using a clever transformation on the latent variables, by converting the constrained latent space to unconstrained space as discussed in Section~\ref{sec:advi}. ADVI models the variational distribution in the unconstrained space as a Gaussian distribution and the transformations applied on the latent variables will satisfy the required constraints on the posterior distribution. The transformation into unconstrained space also mitigates the constraints of support matching that are essential, when choosing a  variational distribution in constraint space, making ADVI a desirable choice for performing variational inference.

 For doing the Bayesian model comparison,~\citet{Krishak_2019}  used nested sampling with the {\tt dynesty}~\citep{dynesty} package  for model comparison, as the {\tt nestle} package was not converging while calculating Bayesian evidence for this problem. To perform model comparison, we calculate PWISE as discussed in Section~\ref{sec:approxevidence},  using  samples from the posterior approximation obtained through ADVI. Table~\ref{tab:cosine_results} shows a comparison of the  results between the proposed approximation to evidence (PWISE) and Nestled Sampling (computed using {\tt dynesty}) for the same sets of priors. We can see that the Bayes factor in both the cases is approximately the same and leads to the same qualitative evidence using Jeffreys scale~\cite{Trotta}. Of course, one caveat in directly applying the Jeffreys scale is that in case the priors for an alternate model are not theoretically motivated, the Jeffreys scale needs to be revised and calibrated to the specific model used~\cite{Gordon}.
  The Bayes factor calculated for H2 compared to  H1 with PWISE is $e^{11.2}$. Hence, we conclude that H2 is favoured over H1, which agrees with the result from ~\citep{Krishak_2019}. For assessing the relative computational cost between both the methods, we executed the nested sampling code given in  ~\cite{Krishak_2019}. The {\tt dynesty} sampling code took about 13 hours (using a single core), whereas ADVI took only 5 minutes, which is two orders smaller than nested sampling.

\begin{table}[t]
    \centering
    \begin{tabular}{|c|c|c|c|c|}
    \hline
    & \multicolumn{2}{c|}{PWISE} & \multicolumn{2}{c|}{dynesty}\\
    \hline
    $H_i$ & ln(D) & Bayes factor & ln(D) & bayes factor \\
    \hline
    $H_1$ & 121.7 & - & 153.7 & - \\
    $H_2$ & 132.9 & $e^{11.2}$ & 168.4 & $e^{14.7}$\\
    \hline
    \end{tabular}
    \caption{Log evidence values and Bayes factor for the two hypotheses computed using PWISE, and {\tt dynesty} packages. This result favors $H_2$, that there is no annual modulation in COSINE-100 data.}
    \label{tab:cosine_results}
\end{table}

\subsection{Exoplanet Discovery Using Radial Velocity Data}

\begin{figure*}[b]
\centering
\includegraphics[width=0.5\textwidth]{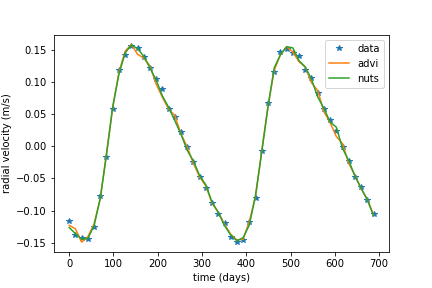}
\includegraphics[width=0.24\textheight]{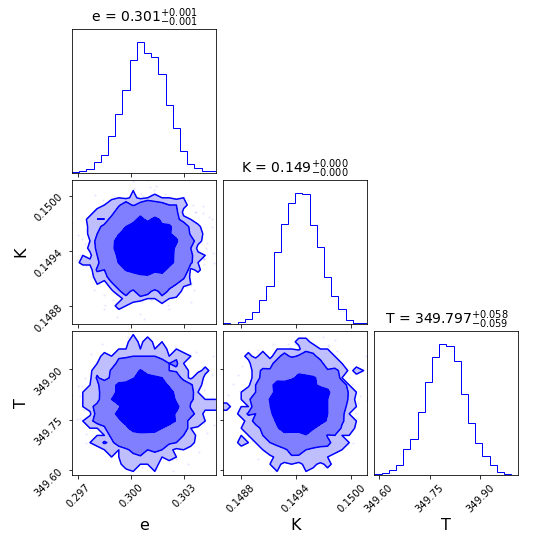}
\caption{Left: Radial velocity as a function of time for a star in a binary system. The orange line is the best fit obtained using ADVI and the green line is obtained from NUTS MCMC. Right: 68\%, 90\% and 95\% credible intervals of parameters obtained using ADVI. The corresponding plots for the same data using MCMC can be found in Fig.~8 of ~\citep{Sharma}.}
\label{fig:rv}
\end{figure*}

The presence of a planet or a companion star results in temporal variations in the radial velocity of the host star. By analyzing the radial velocity data, one can draw inferences about the ratio of masses between the host planet and the companion, and orbital parameters like the period and eccentricity. For this purpose, a MCMC package has been designed  called {\tt Exofit}~\citep{balan2009exofit}, which enables the retrieval of the orbital parameters of exoplanets from radial velocity measurements. We  shall determine  the orbital parameters using both MCMC and ADVI techniques and compare the results.

The first step involves  defining a model and imposing  priors on the latent variables. We follow the model defined in section 2.2 of \cite{balan2009exofit}. The equations used for the analysis are now discussed.
The radial velocity of a star of mass $M$ in a binary system with companion of mass $m$ in an orbit with time period $T$, inclination $I$ and eccentricity $e$ is given by:
\begin{equation}
v(t) = k[\cos(f + \omega) + e\cos \omega] + v_0,
\label{eq:radial_velocity}
\end{equation}
where
\begin{equation}
k = \frac{(2\pi G)^{1/3}m \sin I}{T^{1/3}(M+m)^{2/3}\sqrt[]{1-e^2}}.
\label{eq:k}
\end{equation}
In Eqs.~\eqref{eq:radial_velocity} and ~\eqref{eq:k},  $v_0$ is the mean velocity of the center of mass of the binary system, $T$ is the orbital period of the planet, and $\omega$ is the angle of the pericenter measured from the ascending point.

If $d_i$ is the observed radial velocity data, the likelihood function is given by~\citep{balan2009exofit}:
\begin{equation}
P(D|\theta, M)= A \exp - \left(\sum_{i=1}^N \left[\frac{(d_i-v_i)^2}{2(\sigma_i^2+s^2)}\right]\right),
\label{eq:exoplanetlikelihood}
\end{equation}
where
$A = (2\pi)^{-N/2}\left[ \prod \limits_{i=1}^N (\sigma_i^2+s^2)^{-1/2}\right]$.
Here, $s$ is an additional systematic term, which is estimated by maximizing the likelihood of Eq.~\eqref{eq:exoplanetlikelihood}.
The choice of priors  for each of the above parameters can be found in Table~\ref{tab:exoplanet_priors}. 
{\tt PyMC3} allows us to easily place these priors on model variables and define our model.

\begin{table}[t]
\caption{The assumed prior distribution of various parameters
and their boundaries. It is similar to choice of priors given by \cite{balan2009exofit}. For the parameters marked as Jeffreys prior, the prior used is equal to the reciprocal of the parameter. We note that modified Jeffreys refers to a slight modification of the standard Jeffreys prior, in which additive constants are added, since the lower limits are zero.~\citep{Gregory}}
\begin{center}
 \begin{tabular}{|c|c|}
 \hline
  Parameter & Priors \\
 \specialrule{0.25pt}{0.75pt}{0.75pt}
  $T (days)$ & Jeffreys \\ 
  $k (ms^{-1})$ & Mod. Jeffreys  \\
  $e$ & Uniform \\
  $\omega (\degree)$ & Uniform \\
  $v_0 (ms^{-1})$ & Uniform \\
  $\tau(\degree)$ & Uniform \\
  $s (ms^{-1})$ & Half Normal \\
  \hline
 \end{tabular}
\end{center}
\label{tab:exoplanet_priors}
\end{table}

The data for this purpose has been  obtained from  \cite{Sharma} and the parameter values obtained from both the procedures are shown in Table ~\ref{tab:exoplanet_results}. We find that ADVI converges to a solution in 10 seconds with a mean error of $1.83 \times 10^{-5}$ whereas MCMC took 31 seconds to converge with a mean error of $1.98\times 10^{-5}$. The results and Bayesian credible intervals are shown in Fig.~\ref{fig:rv} and agree with the corresponding results from ~\citep{Sharma}. (cf.  Figure 8 of \citep{Sharma}.) 

\begin{table}[t]
\caption{The parameter values from both MCMC (computing using {\tt PyMC3}) and ADVI for determination of exoplanet parameters from radial velocity data. 
Both of these  are comparable to the actual values obtained from \citep{Sharma}, which are used to generate the synthetic data used for this analysis.}
\begin{center}
 \begin{tabular}{|c|c|c|c|}
 \hline
  Parameter & Actual & MCMC & ADVI\\
  \specialrule{0.25pt}{0.75pt}{0.75pt}
  $T(days)$ & 350 & 349.746 & 349.630 \\ 
  $k(ms^{-1})$ & 0.105 & 0.150 & 0.150  \\
  $e$ & 0.300 &  0.301 & 0.303\\
  $\omega(\degree)$ & -90 & -90.298 & -90.241 \\
  $v0(ms^{-1})$ & 0 & 0.004 & 0.004 \\
  $\tau(\degree)$ & 87.5 & 89.954 & 89.954 \\
  \hline
 \end{tabular}
\end{center}
\label{tab:exoplanet_results}
\end{table}
\subsection{Testing the Periodic $G$ Claim}
\citet{Anderson} have argued for a periodicity of 5.9 years in the CODATA measurements of Newton's gravitational constant $G$, which also show strong correlations with similar variations in the length of the day.
These results have been disputed by \citet{pitkin2015comment} using Bayesian inference  as well as by \citet{Desai} using frequentist analysis, both of which argued that the data for $G$ can be explained without invoking any sinusoidal modulations.
\cite{pitkin2015comment} tested this claim by performing Bayesian model selection using samples generated from MCMC and found from the Bayesian Odds ratio that the data favored a constant value of $G$ with some extra noise over a periodic modulation of $G$ by a factor of $e^{30}$.
We performed model selection using ADVI and {\tt nestle} on the data provided by \citet{pitkin2015comment} to compare the accuracy of variational inference approach.

We compute the Bayesian evidence for all the four hypotheses  considered by Pitkin using the same notation as in~\citet{pitkin2015comment} and compare them as follows:
\begin{enumerate}
    \item  $H_1$ - the data variation can be described by Gaussian noise given by the experimental errors and an unknown offset;
    \item  $H_2$ - the data variation can be described by Gaussian noise given by the experimental errors, an unknown offset and an unknown systematic noise term;
    \item  $H_3$ - the data variation can be described by Gaussian noise given by the experimental errors, and unknown offset, and a sinusoid with unknown period, phase and amplitude;
    \item $H_4$ - the data variation can be described by Gaussian noise given by the experimental errors, an unknown offset, an unknown systematic noise term, and a sinusoid with unknown period, phase and amplitude;
\end{enumerate}
The general model used is
$$
m_i(A, P, \phi_0, T_i, t_0)  = A\sin{(\phi_0 + 2\pi (T_i-t_0)/P)} + \mu_G ,
$$
where $A$ is the  sinusoid amplitude, $P$ is the period, $\phi_0$ is the initial phase, $t_0$ is the initial epoch and $\mu_G$ is an overall offset. The details of the model and assumptions can be found in \citep{pitkin2015comment}.
We have assumed a Gaussian likelihood and uniform prior for all the parameters. Following the model defined by \citep{pitkin2015comment}, we perform model selection using the approximate evidence calculated using the PWISE. Our results computed using PWISE and {\tt nestle} can be found in  Table~\ref{tab:periodic_results} . The log evidence for all the hypotheses are comparable, except for $H_3$. However, even for $H_3$, the Bayes factor (compared to H1) using both the methods qualitatively lead to the same conclusion using Jeffreys scale, viz. $H_3$ been decisively favored over $H_1$. All the experiments were completed under a minute and the time taken by both ADVI and nested sampling are similar.

\begin{table}[t]
    \centering
    \begin{tabular}{|c|c|c|c|c|}
    \hline
    & \multicolumn{2}{c|}{PWISE} & \multicolumn{2}{c|}{nestle}\\
    \hline
    $H_i$ & ln(D) & Bayes factor & ln(D) & Bayes factor \\
    \hline
    $H_{1}$  & 227.5 & - & 232.1 & - \\
    $H_{2}$  & 364.6 & $e^{137.1}$ & 364.7 & $e^{132.6}$ \\
    $H_{3}$ & 243.4 & $e^{15.9}$ & 313.8 & $e^{81.7}$ \\
    $H_{4}$ & 362.9 & $e^{135.4}$ & 364.9 & $e^{132.8}$ \\
    \hline
    \end{tabular}
    \caption{Log evidence values for the four hypotheses and Bayes factor computed with respect to  $H_1$ calculated using both PWISE and {\tt nestle} package. The log evidence for all hypotheses are comparable, except for $H_3$. However, even for $H_3$, the Bayes factor using both the methods qualitatively leads to the same conclusion using Jeffreys scale of $H_3$ been decisively favored over $H_1$.}
    \label{tab:periodic_results}
\end{table}

\subsection{Statistical significance of spectral lag transition in GRB 160625B }
~\citet{Meszaros} have detected  a spectral lag transition in the spectral lag data of GRB 1606025B, which they have argued could be a signature of  the violation of Lorentz invariance (LIV).
\citet{2017APh....94...17G} perform a  frequentist model comparison test to ascertain the statistical significance  of this claim for a transition from positive to negative time lags ,and showed the significance of this detection is  about 3-4$\sigma$, depending on the specific model used for LIV.

\begin{figure*}[b]
\includegraphics[width=0.48\textwidth]{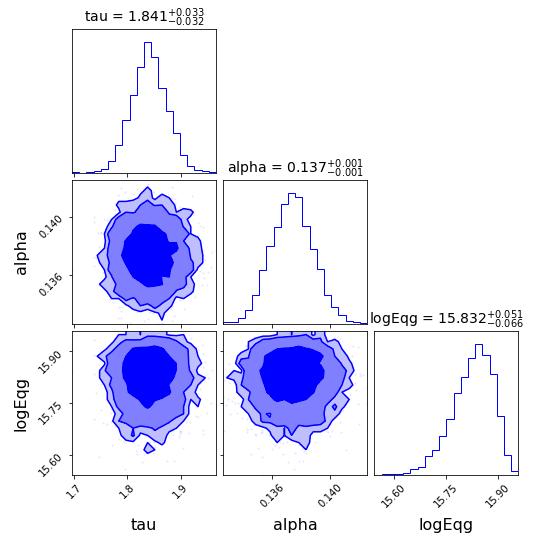}
\hspace*{\fill}
\includegraphics[width=0.48\textwidth]{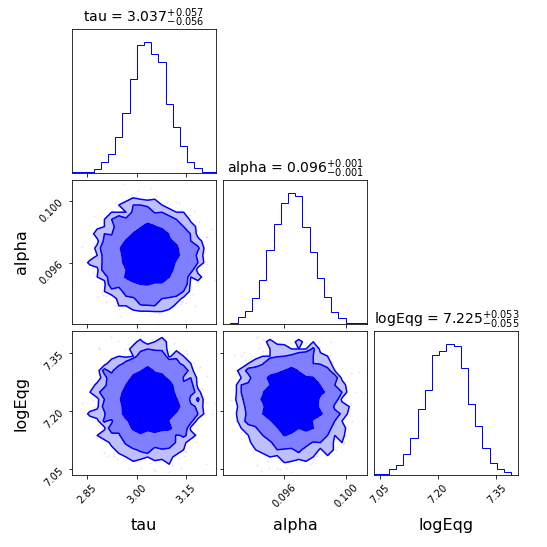}
\caption{Left: ADVI based marginalized credible intervals of the linear (n = 1) LIV fit for the spectral  lag energy data. Right: ADVI based Marginalized parameter constraints of the linear (n = 2) LIV fit  for the spectral lag energy data.  Both the plots were generated using the {\tt corner.py} module~\citep{corner}. The corresponding parameter constraints obtained using MCMC can be found in Figs.~3 and 4 from ~\citep{Meszaros}, and they agree with these contours.}
\label{fig:paramest}
\end{figure*}

For this analysis, ~\citet{Meszaros} have fit these observed lags to a sum of two components: an assumed functional form for the intrinsic time lag due to astrophysical mechanisms and
an energy-dependent speed of light due to quadratic and linear LIV models (See Eqns.~2 and 5 of ~\citet{Meszaros}). 
Using the same equations, we first carry out parameter estimation using ADVI and our best-fit model can be found in Fig.~\ref{fig:paramest}. Again, a Gaussian likelihood and uniform prior was used for this analysis. 

Furthermore, we supplement the studies in \citet{2017APh....94...17G} by  performing Bayesian  model selection using ADVI by fitting a variational family on each of the three models, consisting of the null hypothesis and two Lorentz violation models. We then calculate the approximate evidence using PWISE to perform model selection as defined in Section ~\ref{sec:approxevidence}. The credible intervals for our parameters can be found in Fig.~\ref{fig:paramest}. The log evidence values and the Bayes factors compared to the  null hypothesis are shown in  Table~\ref{tab:LIV} for both Nested sampling (using {\tt nestle}) and PWISE. We see that they are comparable in both the cases and would lead to the same conclusion using Jeffreys scale. For this example, all the experiments were completed under a minute and the time taken by both ADVI and nested sampling are similar.
Using Jeffery's scale we can say that $n=2$ (quadratic) LIV model is significantly favoured by the data over the other two models, which is in agreement with the information theory based model comparisons carried out in ~\citep{2017APh....94...17G}.

\begin{table}[b]
    \centering
    \begin{tabular}{|c|c|c|c|c|}
    \hline
    & \multicolumn{2}{c|}{PWISE} & \multicolumn{2}{c|}{nestle}\\
    \hline
    $H_i$ & ln(D) & bayes factor & ln(D) & bayes factor \\
    \hline
    $H_{n=1}$  & -29.5 & $e^{16.4}$ & -26.9 & $e^{18.6}$ \\
    $H_{n=2}$  & -26.3 & $e^{19.6}$ & -23.9 & $e^{21.6}$\\
    $H_{null}$ & -45.9 & - & -45.5 & - \\ 
    \hline
    \end{tabular}
    \caption{Log Evidence values computed using PWISE and {\tt nestle} package and Bayes factor for hypothesis n=1 and n=2 LIV, when compared to the  null hypothesis are shown.}
    \label{tab:LIV}
\end{table}

\subsection{Estimating the mass of a galaxy cluster with weak lensing}
The propagation of light is affected by the gravitational field it passes through along its way from the observer. This effect is called gravitational lensing~\citep{Falco}. The distortion in the image of an object compared to its true intrinsic shape  is usually known as  weak lensing. \citep{Hoekstra:2013via} outline  how the mass of galaxy clusters  and mass-concentration relation can be obtained using weak lensing. Here,  we use  MCMC to estimate the logarithm of  the virial mass ($\log_{10} M_{200}$) and the concentration parameter $c$ from synthetic lensing observations.

Variational inference and Metropolis-Hastings MCMC were used to calculate  the aforementioned lensing parameters. The dataset used for this analysis was downloaded from  this \href {https://cloud.physik.lmu.de/index.php/s/fac80dcbbccd5258a0d4ae1f9921adb9/download}{this url}. This lensing catalog has been randomly sampled from the shear map of a simulated galaxy cluster using simulations done in~\citep{Becker}, who used mock galaxy clusters from cosmological simulations to study the bias and scatter in mass measurements of clusters. These simulations were created using an Adaptive Refinement Tree~\citep{Kravtsov} based on the cosmological parameters from WMAP7  analysis~\citep{Komatsu}. More details on these simulations and the identification of galaxy cluster halos can be found in ~\citep{Becker}.
A corresponding cookbook for computing the cluster masses using MCMC has also been made available  \href{https://owncloud.physik.uni-muenchen.de/index.php/s/NqqGb1OslXEIy80#pdfviewer}{here}, wherein more details of the equations used can be found, and which we use for reconstructing the mass and concentration parameter. We have used a Gaussian likelihood and uniform priors for the concentration and logarithm of the mass.

\begin{figure*}[t]
\centering
\includegraphics[width=0.48\textwidth,keepaspectratio]{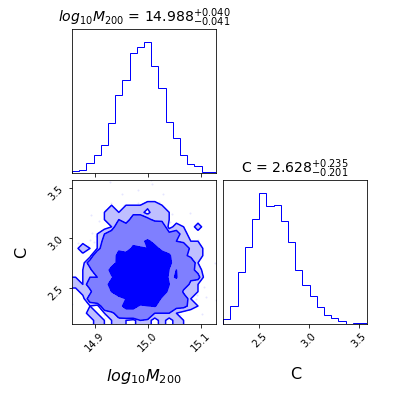}
\hspace*{\fill}
\includegraphics[width=0.48\textwidth,keepaspectratio]{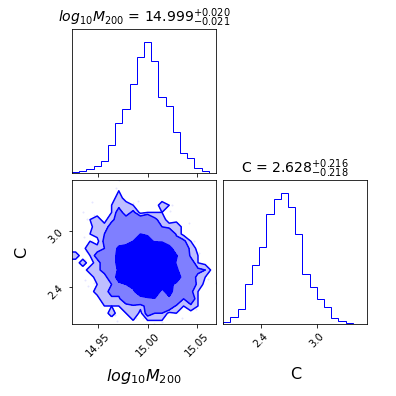}

\caption{Left : Credible intervals for parameter estimates using ADVI. Right: Credible intervals for parameter estimates using {\tt emcee} MCMC sampler. The credible intervals were plotted using {\tt Corner} python module. Note that $M_{200}$ is expressed in terms of $M_{\odot}$.}
\label{fig:wl}
\end{figure*}

For this example, we have used used {\tt pymc3} to run ADVI and {\tt emcee} to run MCMC experiments. MCMC took about 313 minutes of clock time running in multi-threaded mode on 25 cores (corresponding to a total CPU time of $25 \times 313$ minutes or about 5 days), whereas ADVI took only 40 minutes running on a single core. We also note that for this dataset we were unable to run MCMC using {\tt PyMC3}, as it ran out of memory because of the large datasize. The credible intervals for the parameters for both MCMC and ADVI can be found in Fig.~\ref{fig:wl}. The credible intervals using both the techniques are in agreement with each other. 

\section{Conclusions}
\label{sec:conclusions}
In this work, we have introduced variational inference, and outlined how it can be used for Bayesian and frequentist parameter estimation by maximizing the posterior/frequentist likelihood. We have also explained how this method can be used to compute the Bayesian evidence (or marginal likelihood), which is needed for Bayesian model comparison. Variational inference has a strong theoretical foundation and with the rise of probabilistic programming frameworks such as {\tt PyMC3}, and the development of generic Variational Inference methods such as Automatic Differentiable Variational Inference (ADVI), it presents a viable alternative to sampling based approaches such as MCMC. 
We have also introduced a approximation to evidence, called posterior weighted importance sampling for evidence (PWISE) which is used as a proxy for Bayesian evidence (or Marginal likelihood).

ADVI is a `black-box' approach which automates the manual steps required for traditional VI using variable transformation and automatic differentiation techniques. As a proof of principle, we apply ADVI to five problems in astrophysics and gravitation from literature  involving parameter estimation or model comparison.  These include   assessment of significance of annual modulation in COSINE-100 determination of orbital parameters from exoplanet radial velocity data, tests of periodicities in the measurements of $G$, looking for a turnover in spectral lag data from GRB 160625B, and determination of galaxy cluster mass using synthetic weak lensing observations.

The results obtained for both the parameter estimation problem  problem were in agreement with the MCMC  results. For model comparison, both the methods point to the same qualitative conclusion using Jeffreys scale. Furthermore, in many cases, we obtained significant speedup when compared with MCMC methods.
This is especially important when dealing with large datasets and highly complex models as the time required for MCMC approach grows exponentially. On the other hand, variational inference reduces the problem to an optimization problem, which performs very well in these conditions, and hence the computational cost does not scale with data size.   The Markov Chains guarantee producing (asymptotically) exact samples from the target density, but they do not scale very well with large datasets. 
Variational inference therefore provides a viable alternative to MCMC sampling by being significantly faster and given the proper choice of variational distribution, only sacrificing slightly in accuracy. The variational inference algorithm is sensitive to the choice of priors and they can be treated like another hyperparameter. 

These five examples of parameter estimation/model comparison from different domains of astrophysics provide proof of principles demonstration of application of variational inference to astrophysical problems, for which MCMC and nested sampling techniques were previously used.
The codes for all the examples given here  is available at \url{https://github.com/geetakrishna1994/varational-inference}.

\begin{acknowledgements}
Geetakrishnasai Gunapati was supported by DST-ICPS grant. Anirudh Jain  was supported by  the Microsoft summer internship program at IIT Hyderabad.   We would like to thank  Daniel Gruen, Soumya Mohanty, Sanjib Sharma,  and Jochen Weller for useful correspondence and making available to us some of the datasets used in this work.
\end{acknowledgements}
\bibliographystyle{pasa-mnras}
\bibliography{main}
\newpage
\section{Appendix}

We derive the ELBO equation for the cosine-100 problem wrt one parameter 'C'. The variational distribution is isometric Gaussian distribution and uniform priors on all the parameters. The likelihood is Gaussian with a mean given by Eq.~\eqref{eq:event_rates}.
{
\begin{strip}
{\small
\begin{flalign*}
\mathrm{ELBO} &= \mathbb{E}_{q(C,p_0,p_1,A,\omega,t_0)}[\log p(\textbf{D}| C,p_0,p_1,A,\omega,t_0)] - \mathrm{KL} (q(C,p_0,p_1,A,\omega,t_0) | | p(C,p_0,p_1,A,\omega,t_0)) & \\
&= \mathbb{E}_{q(p_0,p_1,A,\omega,t_0)}\bigg[ \mathbb{E}_{q(C)}\bigg[\log p(\textbf{D}| C,p_0,p_1,A,\omega,t_0)] + \log\frac{p(C)}{q(C)}\bigg]\bigg]- \mathrm{KL}\big(q(p_0,p_1,A,\omega,t_0) | | p(p_0,p_1,A,\omega,t_0)\big) \\
\end{flalign*}%
}%
{\small
Gaussian likelihood : 
{\setlength{\abovedisplayskip}{0pt}%
\begin{flalign*}
\log p(\textbf{D}| C,p_0,p_1,A,\omega,t_0) &= \log B - \sum_{i}\bigg(\frac{\big(r_i - \big(C + p_0\exp{(\frac{-\ln{2} \cdot t_i}{p_1})}+A\cos{\omega(t_i-t_0)}\big) \big)^{2}}{2 \sigma_{i}^{2}}\bigg) &\\
&= \log B - \sum_{i}\bigg(\frac{\big((r_i - T_i) - C \big)^{2}}{2 \sigma_{i}^{2}}\bigg)  \\
&= \log B - \sum_{i}\bigg(\frac{(r_i - T_i)^{2}+C^2-2 C(r_i-T_i)}{2 \sigma_{i}^{2}}\bigg) , 
\end{flalign*}%
}%
}%
where $B$ is a normalizing constant for the Gaussian distribution and $T_i = p_0\exp{(\frac{-\ln{2} \cdot t_i}{p_1})}+A\cos{\omega(t_i-t_0)}$.
\\
\\
Prior on C:
{\small
{\setlength{\abovedisplayskip}{6pt}%
\begin{flalign*}
 \log p(C) &= - \log (C_{max} - C_{min}) &
\end{flalign*}%
}%
}%
Variational Distribution of C (Gaussian):
{\small
{\setlength{\abovedisplayskip}{0pt}%
\begin{flalign*}
 \log q(C) &= - \log \sqrt{2\pi\sigma_{C}^{2}} - \frac{(C - \mu_{C})^{2}}{2\sigma_{C}^{2}} & 
\end{flalign*}%
}%
}%
{\small
{\setlength{\abovedisplayskip}{0pt}%
\begin{flalign*}
 \log p(\textbf{D}| C,p_0,p_1,A,\omega,t_0) + \log p(C) - \log q(C) &=  \log B - \sum_{i}\bigg(\frac{(r_{i}-T_i)^{2}+C^2-2 C(r_i-T_i)}{2 \sigma_{i}^{2}}\bigg) - \log (C_{max} - C_{min}) &\\ 
&\quad +  \log \sqrt{2\pi\sigma_{C}^{2}} + \frac{(C - \mu_{C})^{2}}{2\sigma_{C}^{2}} \\
&= \log \frac{B\sqrt{2\pi\sigma_{C}^{2}}}{C_{max}-C_{min}} + C^{2}\bigg(\frac{1}{2\sigma_{C}^{2}} - \sum_i\frac{1}{2\sigma_{i}^{2}}  \bigg)
+C\bigg(\sum_{i}\frac{r_i-T_i}{\sigma_i^2} - \frac{\mu_C}{\sigma_C^2}\bigg)\\
&\quad +\bigg(\frac{\mu_C^2}{2\sigma_C^2} - \sum_{i}\frac{(r_i-T_i)^2}{2\sigma_i^2}\bigg)
\end{flalign*}%
}%
}%
We can evaluate the expectation of the above term wrt $q(C)$.
{\small
{\setlength{\abovedisplayskip}{0pt}%
\begin{flalign*}
 \mathbb{E}_{q(C)}\bigg[\log p(\textbf{D}| C,p_0,p_1,A,\omega,t_0)] + \log(p(C)) - \log(q(C))\bigg] &= \log \frac{B\sqrt{2\pi\sigma_{C}^{2}}}{C_{max}-C_{min}} +\bigg(\frac{\mu_C^2}{2\sigma_C^2} - \sum_{i}\frac{(r_i-T_i)^2}{2\sigma_i^2}\bigg) & \\
& \quad  + \bigg(\frac{1}{2\sigma_{C}^{2}} - \sum_i\frac{1}{2\sigma_{i}^{2}}  \bigg) \mathbb{E}_{q(C)}[C^2] + \bigg(\sum_{i}\frac{r_i-T_i}{\sigma_i^2} - \frac{\mu_C}{\sigma_C^2}\bigg) \mathbb{E}_{q(C)}[C]  \\
&= \log \frac{B\sqrt{2\pi\sigma_{C}^{2}}}{C_{max}-C_{min}} +\bigg(\frac{\mu_C^2}{2\sigma_C^2} - \sum_{i}\frac{(r_i-T_i)^2}{2\sigma_i^2}\bigg)  \\
& \quad + \bigg(\frac{1}{2\sigma_{C}^{2}} - \sum_i\frac{1}{2\sigma_{i}^{2}}  \bigg)(\sigma_C^2+\mu_C^2) + \bigg(\sum_{i}\frac{r_i-T_i}{\sigma_i^2} - \frac{\mu_C}{\sigma_C^2}\bigg)(\mu_C) \\
&= \frac{1}{2} + \log \frac{B\sqrt{2\pi\sigma_{C}^{2}}}{C_{max}-C_{min}}
-\sum_i \frac{1}{2\sigma_i^2}\bigg((r_i-T_i-\mu_C)^2 + \sigma_C^2\bigg)
\end{flalign*}%
}%
}%
{\footnotesize
{\setlength{\abovedisplayskip}{0pt}%
\begin{flalign*}
\mathrm{ELBO} &=  \mathbb{E}_{q(p_0,p_1,A,\omega,t_0)}\bigg[\frac{1}{2} + \log \frac{B\sqrt{2\pi\sigma_{C}^{2}}}{C_{max}-C_{min}}
-\sum_i \frac{1}{2\sigma_i^2}\bigg((r_i-T_i-\mu_C)^2 + \sigma_C^2\bigg) \bigg]- \mathrm{KL}\big(q(p_0,p_1,A,\omega,t_0) | | p(p_0,p_1,A,\omega,t_0)\big) & \\
&= \frac{1}{2} + \log \frac{B\sqrt{2\pi\sigma_{C}^{2}}}{C_{max}-C_{min}} -\sum_i \frac{1}{2\sigma_i^2}\bigg(\mathbb{E}_{q(p_0,p_1,A,\omega,t_0)}\bigg[(r_i-T_i-\mu_C)^2\bigg]+ \sigma_C^2\bigg)- \mathrm{KL}\big(q(p_0,p_1,A,\omega,t_0) | | p(p_0,p_1,A,\omega,t_0)\big)
\end{flalign*}%
}%
}%
\end{strip}
}%
\end{document}